\documentclass[aps,prl,twocolumn,showpacs,superscriptaddress,groupedaddress,notitlepage]{revtex4-1}
\usepackage[english]{babel}
\usepackage{amsmath,amssymb}
\usepackage{graphicx}
\usepackage{dcolumn}
\usepackage{bm}
\usepackage{amsmath}
\usepackage{xcolor}
\usepackage[applemac]{inputenc}
\usepackage{mathtools}
\usepackage{accents}
\usepackage[labelformat=empty,caption=false]{subfig}
\usepackage{keyval}
\newlength{\dhatheight}

\newcommand{\average}[1]{\left\langle #1 \right\rangle}

\newcommand*{\diff}{\mathop{}\!\mathrm{d}}
\renewcommand{\phi}{\varphi}

\newcommand{\derpar}[2]{\frac{\partial #1}{\partial #2}}

\newcommand{\dersecpar}[2]{\frac{\partial^2 #1}{\partial #2^2}}
\newcommand{\dermixpar}[3]{\frac{\partial^2 #1}{\partial #2 \, \partial #3}}

\newcommand{\MittagLeffler}[3]{E_{#1,#2}(#3)}

\begin{document}

\title{Anomalous processes with general waiting times: functionals and multi-point structure}
\author{Andrea Cairoli}
\affiliation{School of Mathematical Sciences, Queen Mary, University of London, 327 Mile End Road, E1 4NS, UK}
\author{Adrian Baule}
\email{Correspondence to: a.baule@qmul.ac.uk}
\affiliation{School of Mathematical Sciences, Queen Mary, University of London, 327 Mile End Road, E1 4NS, UK}
\date{\today}
\begin{abstract}
Many transport processes in nature exhibit anomalous diffusive properties with non-trivial scaling of the mean square displacement, e.g., diffusion of cells or of biomolecules inside the cell nucleus, where typically a crossover between different scaling regimes appears over time. Here, we investigate a class of anomalous diffusion processes that is able to capture such complex dynamics by virtue of a general waiting time distribution. We obtain a complete characterization of such generalized anomalous processes, including their functionals and multi-point structure, using a representation in terms of a normal diffusive process plus a stochastic time change. In particular, we derive analytical closed form expressions for the two-point correlation functions, which can be readily compared with experimental data.
\end{abstract}


\maketitle

Diffusive transport is usually classified in terms of the mean square displacement (MSD): $MSD(t)=\average{(R(t)-R_0)^2}$, where $R(t)$ is a time-dependent stochastic vector, either the position or the velocity, and $R_0$ is its initial value. Motivated by numerous experimental results of the last decade (see \cite{metzler2000random,hofling2013anomalous} and references therein for an up-to-date overview), we usually distinguish between normal and anomalous diffusive processes for which the MSD scales linearly in time or as a power-law respectively \cite{klafter1996beyond,metzler2000random,klages2008anomalous,sokolov2012models,hofling2013anomalous,zaburdaev2015Levy}. However, due to the improvement of experimental techniques, evidence of more complicated nonlinear MSDs, characterized by different scaling regimes over the measurement time, have been found in recent experiments of diffusion in biophysical systems\cite{selmeczi2005cell,selmeczi2008cell,dieterich2008anomalous,campos2010persistent,harris2012generalized,caspi2000enhanced,levi2005chromatin,brangwynne2007force,bronstein2009transient,bruno2009transition,senning2010actin,jeon2012anomalous,weber2012nonthermal,von2013anomalous,tabei2013intracellular,javer2014persistent}.
Here, we investigate a general class of anomalous diffusive processes that can capture such complicated MSD behaviour by means of a generalized waiting time distribution. We provide a complete characterization of these processes including: (i) The stochastic description of the microscopic diffusive dynamics; (ii) Evolution equations for the probability density function (pdf) of the process and its associated time-integrated observables; (iii) The multi-point correlation functions. Our general model includes as a special case the continuous time random walk (CTRW), which is widely used to model MSDs with a power-law scaling \cite{bronstein2009transient,jeon2012anomalous,tabei2013intracellular}. Even though CTRWs have been in the focus of theoretical research on anomalous diffusion for almost two decades, the full characterization of CTRWs in terms of (i)--(iii) has not been presented so far. In this letter we provide the key for such a complete stochastic description by using the stochastic calculus of random time changes.


The stochastic trajectory of a CTRW $Y(t)$ is expressed in terms of coupled Langevin equations \cite{fogedby1994langevin}:
\begin{subequations}
\begin{align}
\dot{X}(s)&=F(X(s))+\sigma(X(s))\cdot\xi(s), \label{eq:1.1a} \\
\dot{T}(s)&=\eta(s), \label{eq:1.1b}
\end{align}
\end{subequations}
where for convenience we focus on a process in 1d. The CTRW is then given by $Y(t)=X(S(t))$, where the process $S$ is defined as the inverse of $T$, or more precisely as the collection of first passage times:
\begin{equation}
S(t)=\inf_{s>0}{\left\{s:T(s)>t\right\}}.
\label{eq:1.1c}
\end{equation} 
The dynamics of $X$ is that of a normal diffusive process in the operational time $s$. Thus, $F(x)$ and $\sigma(x)$ satisfy standard conditions \cite{revuz1999continuous} and we adopt the It\^o convention for the multiplicative term of Eq.~\eqref{eq:1.1a}. We also require $\xi(s)$ to represent white Gaussian noise with $\average{\xi(s)}=0$ and $\average{\xi(s_1)\xi(s_2)}=\delta(s_2-s_1)$. The noise $\eta(s)$ models the waiting times of the anomalous diffusion process in the operational time $s$, which we assume independent from the $X$ process, i.e., the noises $\xi(s)$ and $\eta(s)$ are statistically independent. In the case of a CTRW, $\eta(s)$ is one-sided stable L\'evy noise of order $0 < \alpha \leq 1$ \cite{cont1975financial} leading to the characteristic function of $T$: $\left<e^{-\lambda T(s)}\right>=e^{-s\lambda^\alpha}$. In physical terms, $T$ represents the elapsed physical time of the process.

In order to extend this picture to arbitrary waiting times, we consider $\eta(s)$ as a more general type of noise, which can be modelled by a general one-sided L\'evy process with finite variation \cite{cont1975financial,applebaum2009Levy}. Such a process satisfies the minimal assumptions needed to assure independent and stationary waiting times and causality of $T$. A complete characterization of $\eta(s)$ is given by its characteristic functional:
\begin{equation}
G[k(s)]=\left\langle e^{\,-\int_{0}^{+\infty}k(s)\eta(s)\diff{s}}\right\rangle=e^{\,-\int_{0}^{+\infty}\Phi(k(s))\diff{s}}.
\label{eq:1.2}
\end{equation} 
Here, the so-called Laplace exponent $\Phi$ is a non-negative function with $\Phi(0)=0$ and monotonically decreasing first derivative. Different functional forms of $\Phi$ correspond to different distribution laws of the waiting times and, consequently, of the renewal process $T$. The renewal nature of $T$ is expressed by Eq.~(\ref{eq:1.1b}) as $T(s)=\int_0^s\eta(s')\diff{s'}$. The characteristic function $\left<e^{-\lambda T(s)}\right>$ is thus directly obtained from Eq.~(\ref{eq:1.2}) by setting $k(s')=\lambda\Theta(s-s')$ leading to $\left<e^{-\lambda T(s)}\right>=e^{-s\Phi(\lambda)}$. Clearly, this implies that $T$ is a sum over waiting time increments $\Delta t=\int_0^{\Delta s}\eta(s')\diff{s'}$ over a small time step $\Delta s$ with characteristic function $\left<e^{-\lambda \Delta t}\right>=e^{-\Delta s\Phi(\lambda)}$, which can be used to simulate the process $Y(t)$ within a suitable discretization scheme \cite{kleinhans2007continuous}. Remarkably, the full multi-point statistics of $T$ becomes easily accessible, because the functional Eq.~(\ref{eq:1.2}) contains the information about the whole noise trajectory. By choosing $\Phi$ suitably, many different waiting time statistics can be captured, i.e., $Y(t)$ can be modelled according to the observed experimental dynamics.
If we choose a power law $\Phi(\lambda)=\lambda^{\alpha}$ we recover the CTRW case with waiting times characterized by a diverging first moment. If instead $\Phi(\lambda)=\lambda$, $T$ is simply a deterministic drift, $T=s$, and $Y(t)$ reduces to a normal diffusion (Brownian limit) with waiting times following an exponential distribution \cite{metzler2000random}.

Along with the properties of $Y(t)$, we also study those of its time-integrated observables, which are naturally defined as functionals of the stochastic trajectory \cite{majumdar2005brownian}:
\begin{equation}
W(t)=\int_0^{t} U(Y(r)) \diff{r}, \label{eq:0.1}
\end{equation}
where $U(x)$ is a smooth integrable function. Clearly, if $Y(t)$ is a velocity and $U(x)=x$, $W(t)$ is the corresponding position. When $Y(t)$ is a normal diffusion the joint pdf $P(w,y,t)=\left\langle \delta(w-W(t))\delta(y-Y(t))\right\rangle$ is provided by the celebrated Feynman-Kac (FK) formula \cite{majumdar2005brownian}. The FK theory provides the crucial connection between the stochastic description of the process in the diffusive limit and the evolution equation for the pdf. When $Y(t)$ is anomalous instead, the computation of the joint pdf reveals profound challenges. In the CTRW case the problem could only be addressed so far using a time discretized description in terms of Master equations leading to a fractional FK equation with a fractional substantial derivative \cite{turgeman2009fractional,carmi2010distributions}. In the following, we provide the missing link between the diffusive description of CTRWs and the fractional FK equation and generalize the connection to arbitrary waiting times.

The monotonicity of $T$ and $S$ implies \cite{baule2005joint}: 
\begin{equation}
\Theta(s-S(t))=1-\Theta(t-T(s)), 
\label{eq:1.3}
\end{equation}
which, together with the continuity of the paths of $S(t)$ and the corresponding It{\^o} formula, provides the relation: 
\begin{equation}
\delta(t-T(s))=\delta(s-S(t))\dot{S}(t), \label{eq:1.4}
\end{equation}
Formally, this equation and the following ones in which derivatives of $S(t)$ appear are to be interpreted in their corresponding integral forms, with $\dot{S}(t)=\lim_{\Delta t \to 0}\frac{S(t+\Delta t)-S(t)}{\Delta t}$ being a shorthand notation for the stochastic increment of the time-change. We can then describe the time-changed process $Y(t)$ and its functional $W(t)$ through the time-changed Langevin equations \cite{kobayashi2011stochastic}: 
\begin{subequations}
\begin{align}
\dot{Y}(t)&=F(Y(t))\dot{S}(t)+\sigma(Y(t))\cdot\xi(S(t))\dot{S}(t), \label{eq:1.5a} \\
\dot{W}(t)&=U(Y(t)) \label{eq:1.5b}
\end{align}
\end{subequations} 

The FK formula describes the time evolution of the Fourier transform of $P(w,y,t)$: $\widehat{P}(p,y,t)=\average{e^{\,i\,p\,W(t)}\delta(y-Y(t))}$ \cite{majumdar2005brownian}, with $\average{\ldots}$ being an average over the realizations of both the noises $\xi(s)$ and $\eta(s)$ in the anomalous case. In the following, $\widehat{g}(k)=\int_0^{+\infty}e^{i k x}g(x)\diff{x}$ denotes the Fourier transform of $g(x)$ and $\widetilde{f}(\lambda)=\int_0^{+\infty}e^{- \lambda t}f(t)\diff{t}$ the Laplace transform of $f(t)$. Our derivation begins with the It\^o formula for the joint time-changed process $Z(t)=(Y(t),W(t))$ \cite{jacod714calcul}:   
\begin{align}
\label{eq:2.1}
&f(Z(t))=f(Z_0)+\frac{1}{2}\int_0^t\dermixpar{f}{y}{w}(Z(t))\diff{[Y,W]_{t}}\\
&+\int_0^t\derpar{}{y}f(Z(t))\diff{Y(t)}+\int_0^t\derpar{}{w}f(Z(t))\diff{W(t)} \notag\\
&+\frac{1}{2}\!\int_0^t\dersecpar{}{y}f(Z(t))\diff{[Y,Y]_{t}}\!+\!\frac{1}{2}\!\int_0^t\dersecpar{}{w}f(Z(t))\diff{[W,W]_{t}} \notag
\end{align}
where the square brackets denote the quadratic variation of two processes \cite{revuz1999continuous}. By using the time-discretized form of Eqs.~(\ref{eq:1.5a}-\ref{eq:1.5b}), the exact relation $[Y,Y]_{t}=\int_{0}^{t}\sigma^2(Y(\tau))\dot{S}(\tau)\diff{\tau}$ \cite{magdziarz2010path,kobayashi2011stochastic} and the fact that $[W,W]_t=0=[Y,W]_t$, we obtain: 
\begin{align}
f(Z(t))=&f(Z_0)+\int_{0}^{t}\derpar{}{w}f(Z(\tau))U(Y(\tau))\diff{\tau} \notag\\
&+\int_{0}^{t}\derpar{}{y}f(Z(\tau))F(Y(\tau))\dot{S}(\tau)\diff{\tau} \notag\\
&+\frac{1}{2}\int_{0}^{t}\dersecpar{}{y}f(Z(\tau))\sigma^{2}(Y(\tau))\dot{S}(\tau)\diff{\tau} \notag\\
&+\int_{0}^{t}\derpar{}{y}f(Z(\tau))\sigma(Y(\tau))\xi(S(\tau))\dot{S}(\tau)\diff{\tau}.
\label{eq:2.2}
\end{align}
If we now evaluate Eq.~\eqref{eq:2.2} for $f(Z(t))=e^{i\,k\,Y(t)+i\,p\,W(t)}$ and take its ensemble average, we derive an equation for $\skew{3.5}\widehat{\widehat{P}}(p,k,t)$. We remark that the last integral in the rhs of Eq.~\eqref{eq:2.2} disappears due to the independence of the increments of $\xi(s)$. Indeed, if we make the inverse transform and recall the Fokker-Planck operator of Eq.~\eqref{eq:1.1a}: $\mathcal{L}_{FP}(y)=-\derpar{}{y}F(y)+\frac{1}{2}\dersecpar{}{y}\sigma^2(y)$, we obtain the equation:  
\begin{multline}
\derpar{}{t}\widehat{P}(p,y,t)=i\,p\,U(y)\,\widehat{P}(p,y,t) \\
+\mathcal{L}_{FP}(y)\derpar{}{t}\left\langle \int_{0}^{t}e^{i\,p\,W(\tau)}\delta(y-Y(\tau))\dot{S}(\tau)\diff{\tau}\right\rangle. 
\label{eq:2.3}
\end{multline}

We still need to relate the expression in brackets in Eq.~\eqref{eq:2.3} to $\widehat{P}(p,y,t)$. We start by manipulating directly $\widehat{P}(p,y,t)$. First, we make the change of variables $\tau=S(r)$ in Eq.~\eqref{eq:0.1}, so that we can write:
\begin{equation}
W(t)=W_{s}(S(t)) \,\,\, , \,\,\,  W_s(s)=\int_0^{s} U(X(\tau)) \eta(\tau)\diff{\tau},
\label{eq:A1}
\end{equation}
where the noise $\eta(s)$ explicitly appears because $r=T(\tau)$ via Eq.~\eqref{eq:1.1c}. We then use $1=\int_0^{+\infty}\delta(s-S(t))\diff{s}$ to write:     
\begin{equation}
\widehat{P}(p,y,t)=\average{\int_0^{+\infty}e^{i\,p\,W_{s}(s)}\delta(y-X(s))\,\delta(s-S(t))\diff{s}}.
\label{eq:A2}
\end{equation}
As a third step, we compute the Laplace transform of Eq.~\eqref{eq:A2}. By using Eq.~\eqref{eq:1.3}, we find the relation: $\int_0^{+\infty}\delta(s-S(t))e^{-\lambda\,t}\diff{t}=\eta(s)e^{-\lambda\,T(s)}$, which provides:  
\begin{equation}
\label{eq:2.4}
\skew{3.5}\widehat{\widetilde{P}}(p,y,\lambda)\!=\!\int_0^{+\infty} \! \! \average{e^{-\lambda\,T(s)+i\,p\,W_s(s)}\eta(s)\delta(y-X(s))}\diff{s}. 
\end{equation}   
This can be further simplified by expressing the $\eta(s)$-dependent part of the integrand as a derivative of the characteristic functional $G[k(l)]$ with $k(l)=\left(\lambda-i\,p\,U(X(l))\right)\Theta(s-l)$. By performing the average with respect to $\eta(s)$ first and using Eq.~\eqref{eq:1.2}, we derive:  
\begin{multline}
\skew{3.5}\widehat{\widetilde{P}}(p,y,\lambda)=\frac{\Phi\left[\lambda - i\,p\,U(y)\right]}{\lambda - i\,p\,U(y)}\times\\
\times  \average{\int_0^{+\infty}e^{-\lambda T(s) + i\,p\,W_s(s)}\delta(y-X(s))\diff{s}}.
\label{eq:2.5}
\end{multline}

Now we need to show that the two integrals in brackets in Eqs.~(\ref{eq:2.3},\ref{eq:2.5}) coincide. Indeed, the inverse Laplace transform of the integral in Eq.~(\ref{eq:2.5}) is given by: $\int_0^{+\infty}e^{i\,p\,W_{s}(s)}\delta(y-X(s))\delta(t-T(s))\diff{s}$. This can then be cast into the integral of Eq.~\eqref{eq:2.3} by using Eq.~\eqref{eq:1.4} and the continuity of the paths of $S$, which implies that no jump terms appear in the expansion of the stochastic integral in Eq.~\eqref{eq:2.3} as a discrete sum. Consequently, we can use Eq.~\eqref{eq:2.5} in Eq.~\eqref{eq:2.3} to write the generalized FK formula: 
\begin{align}
\label{eq:2.6}
&\derpar{}{t}\widehat{P}(p,y,t)=i\,p\,U(y)\,\widehat{P}(p,y,t)+\mathcal{L}_{FP}(y) \times \\
&\times \left[\derpar{}{t}-i\,p\,U(y)\,\right]\int_0^{t}\,K(t-\tau)\,e^{i\,p\,U(y)(t-\tau)}\,\widehat{P}(p,y,\tau)\diff{\tau}, \notag
\end{align}
where the memory kernel is related to $\Phi$ by: 
\begin{equation}
\widetilde{K}(\lambda)=\Phi(\lambda)^{-1}.
\label{eq:2.9}
\end{equation}
If we set $p=0$, we obtain a generalized Fokker-Planck equation for $P(y,t)=\average{\delta(y-Y(t))}$: 
\begin{equation}
\derpar{}{t}P(y,t)=\mathcal{L}_{FP}(y)\derpar{}{t}\int_0^{t}\,K(t-\tau)\,P(y,\tau)\diff{\tau}. 
\label{eq:2.7}
\end{equation}  
In the special case where $W(t)$ corresponds to the position, i.e., $U(x)=x$, Eq.~(\ref{eq:2.6}) yields a generalized Klein-Kramers equation exhibiting retardation effects \cite{Friedrich2006Anomalous,Friedrich2006Exact}. The Brownian limit is achieved for $K(t)=1$, where Eqs.~(\ref{eq:2.6},\ref{eq:2.7}) reproduce the standard FK formula \cite{majumdar2005brownian}, as well as the Fokker-Planck and Klein-Kramers equations \cite{risken1989fokker}, respectively. In the CTRW case, $\Phi(\lambda)=\lambda^\alpha$ so that the integral operators in Eqs.~(\ref{eq:2.6},\ref{eq:2.7}) coincide with the fractional substantial derivative \cite{Friedrich2006Anomalous,Friedrich2006Exact} and the standard Riemann-Liouville operator respectively, thus recovering the well-known CTRW results \cite{metzler1999deriving,Friedrich2006Anomalous,Friedrich2006Exact,magdziarz2008equivalence,turgeman2009fractional,Orzel2011FKKE}. We highlight that this derivation of Eq.~(\ref{eq:2.6}) provides the generalization of the Feynman-Kac theorem to anomalous processes with arbitrary waiting time distributions.

We now focus on deriving the multi-point statistics of $Y(t)$ and $W(t)$.
Arbitrary two-point functions of $Y(t)$ are expressed as
\begin{align}
\label{eq:3.1}
&\average{f(Y(t_1),Y(t_2))} \\
&=\!\int_0^{+\infty}\!\!\!\int_0^{+\infty}\!\!\average{f(X(s_2),X(s_1))}h(s_2,t_2;s_1,t_1)\diff{s_2}\diff{s_1} \notag
\end{align}
where $h(s_2,t_2;s_1,t_1)=\average{\delta(s_2-S(t_2))\delta(s_1-S(t_1))}$ is the two-point pdf of $S(t)$. Using Eq.~\eqref{eq:1.3} we obtain: $h(s_2,t_2;s_1,t_1)=\dermixpar{}{s_2}{s_1}\average{\Theta(t_2-T(s_2))\Theta(t_1-T(s_1))}$. Therefore, the Laplace transform of $h$ is related to the two-point characteristic function $Z(\lambda_2,s_2;\lambda_1,s_1)=\average{e^{-\lambda_2 T(s_2)}e^{-\lambda_1 T(s_1)}}$:
\begin{equation}
\widetilde{h}(s_2,\lambda_2;s_1,\lambda_1)=\frac{1}{\lambda_1\lambda_2}\dermixpar{}{s_2}{s_1}Z(\lambda_2,s_2;\lambda_1,s_1).
\label{eq:3.2}
\end{equation}
The computation of $Z$ follows straightforwardly by distinguishing the two cases $t_2>t_1$ and $t_2<t_1$ and by recalling the independence of the increments of $T(s)$:
\begin{multline}
Z(\lambda_2,s_2;\lambda_1,s_1)=\Theta(s_2-s_1)e^{-s_1\,\Phi\left(\lambda_1+\lambda_2\right)}e^{-\left(s_2-s_1\right)\,\Phi\left(\lambda_2\right)} \\
+\Theta(s_1-s_2)e^{-s_2\,\Phi\left(\lambda_1+\lambda_2\right)}e^{-\left(s_1-s_2\right)\,\Phi\left(\lambda_1\right)}. 
\label{eq:3.3}
\end{multline}
This result can then be substituted in Eq.~\eqref{eq:3.2} to derive:
\begin{align}
&\widetilde{h}(s_2,\lambda_2;s_1,\lambda_1)= \notag\\
&\quad \delta(s_2-s_1)\frac{\Phi(\lambda_1)-\Phi(\lambda_1+\lambda_2)+\Phi(\lambda_2)}{\lambda_1\lambda_2}e^{-s_1\Phi(\lambda_1+\lambda_2)} \notag\\
&\quad +\Theta(s_2-s_1)\frac{\Phi(\lambda_2)\left[\Phi(\lambda_1+\lambda_2)-\Phi(\lambda_2)\right]}{\lambda_1\lambda_2}\times \notag\\
&\quad\quad \times e^{-s_1\Phi(\lambda_1+\lambda_2)}e^{-(s_2-s_1)\Phi(\lambda_2)} \notag\\
&\quad +\Theta(s_1-s_2)\frac{\Phi(\lambda_1)\left[\Phi(\lambda_1+\lambda_2)-\Phi(\lambda_1)\right]}{\lambda_1\lambda_2}\times \notag\\
&\quad\quad \times e^{-s_2\Phi(\lambda_1+\lambda_2)}e^{-(s_1-s_2)\Phi(\lambda_1)}.
\label{eq:3.4}
\end{align} 
With the explicit expression for $\widetilde{h}$ the average $\average{f(Y(t_2),Y(t_1))}$ can be calculated for arbitrary $X$ dynamics. Moreover, also correlation functions of $W(t)$ can be easily derived using: 
\begin{align}
\label{eq:3.5}
&\average{\widetilde{W}(\lambda_1) \widetilde{W}(\lambda_2)}=\frac{1}{\lambda_1 \lambda_2}\times \\
&\quad\!\!\!\! \times\!\! \int_0^{+\infty}\!\!\!\!\int_0^{+\infty}\!\!\!\!\average{U(X(s_2)),U(X(s_1))}\!\widetilde{h}(s_2,\lambda_2;s_1,\lambda_1)\!\diff{s_2}\!\diff{s_1}. \notag
\end{align}
The corresponding formulas for higher orders are derived in full analogy, thus providing access to the complete multi-point structure of $Y(t)$ and $W(t)$.
If $X(s)$ has the usual form $\average{X(s_1)X(s_2)}=H(s_2-s_1)$ in the stationary regime, with $H$ being a smooth decreasing function, the inverse transform of Eq.~\eqref{eq:3.1} can be made explicitly:
\begin{equation}
\average{Y(t_1)Y(t_2)}=f_1(t_2)+\int_0^{t_1} K(\tau)f_2(t_2-\tau) \diff{\tau},
\label{eq:3.6}
\end{equation}
where the transient terms can be shown to disappear for $t_2>t_1\to\infty$ using Tauberian theorems. Here, we define the functions: $\widetilde{f_{1}}(\lambda)=\frac{\Phi(\lambda)}{\lambda}\widetilde{H}(\Phi(\lambda))$ and $\widetilde{f_{2}}(\lambda)=\frac{[\Phi(\lambda)]^2}{\lambda}\widetilde{H}(\Phi(\lambda))$. The specific form of these functions can be obtained once both $X$ and $\Phi$ are specified. Eq.~(\ref{eq:3.6}) is remarkable because the two-point function is expressed in terms of an integral of \textit{single-time} functions, highlighting a simple underlying structure.

\begin{figure}[!htb]
\subfloat[][]{\includegraphics[scale=0.135]{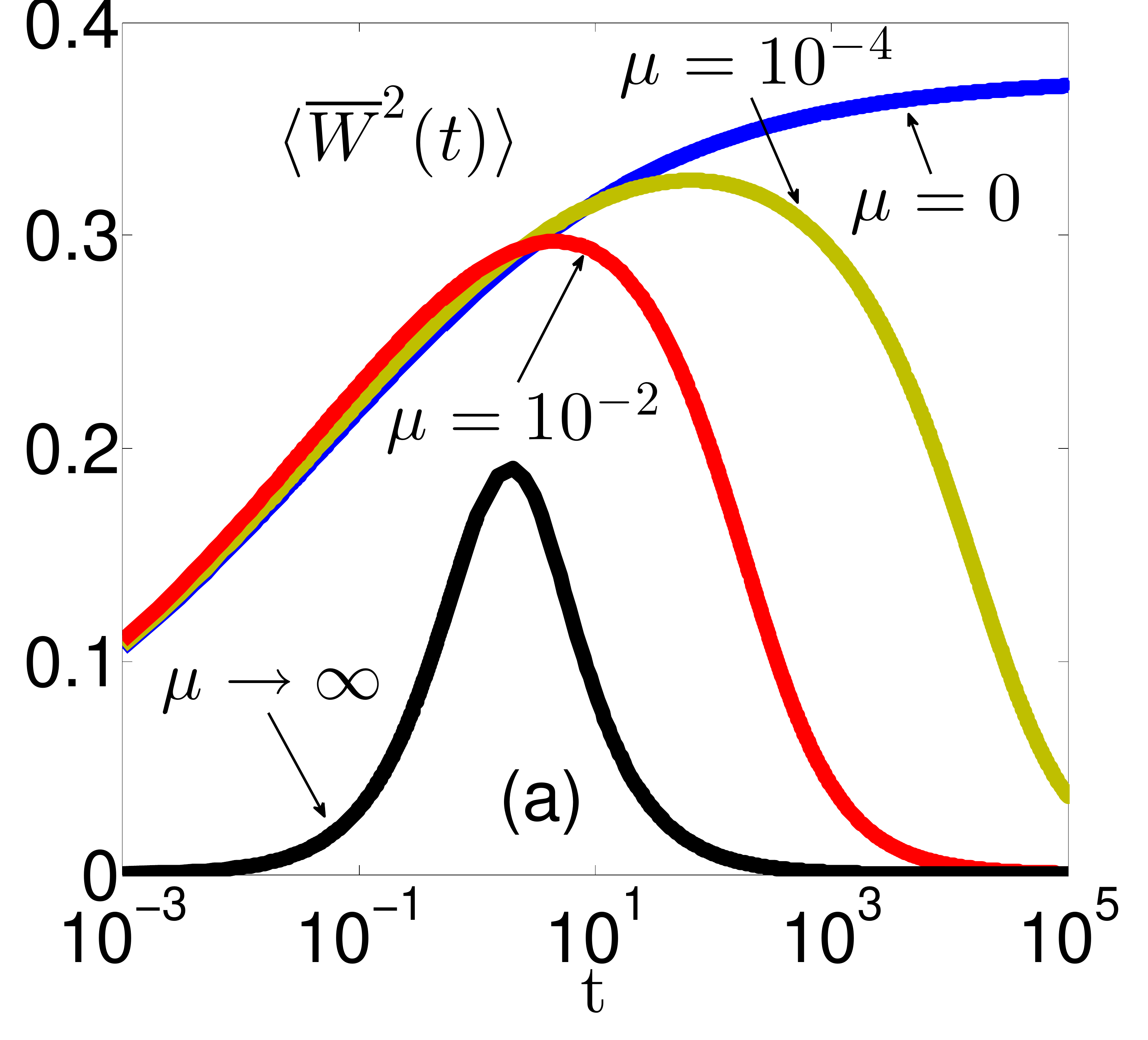}}
\subfloat[][]{\includegraphics[scale=0.135]{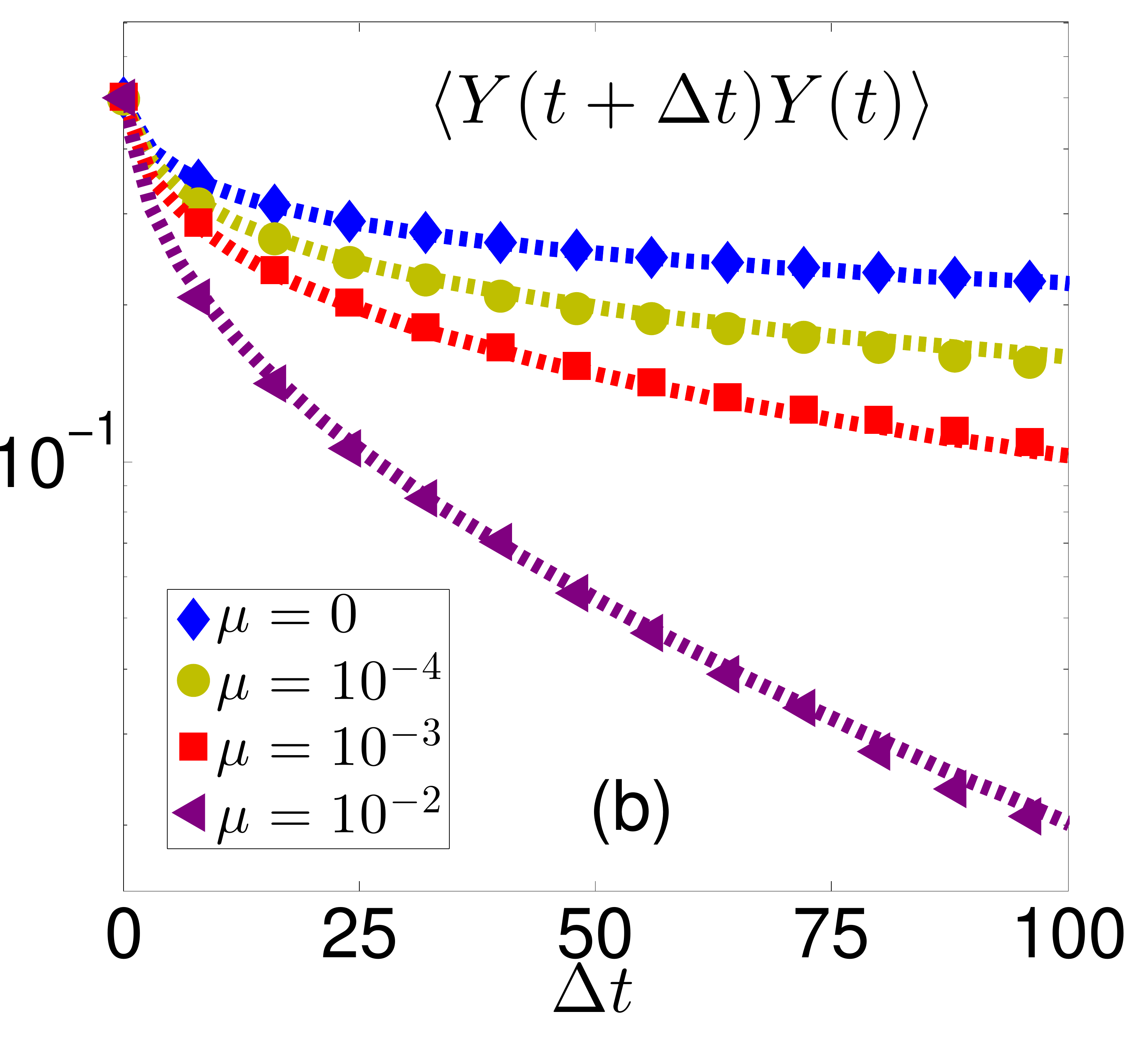}}\\[-7ex]
\subfloat[][]{\includegraphics[scale=0.135]{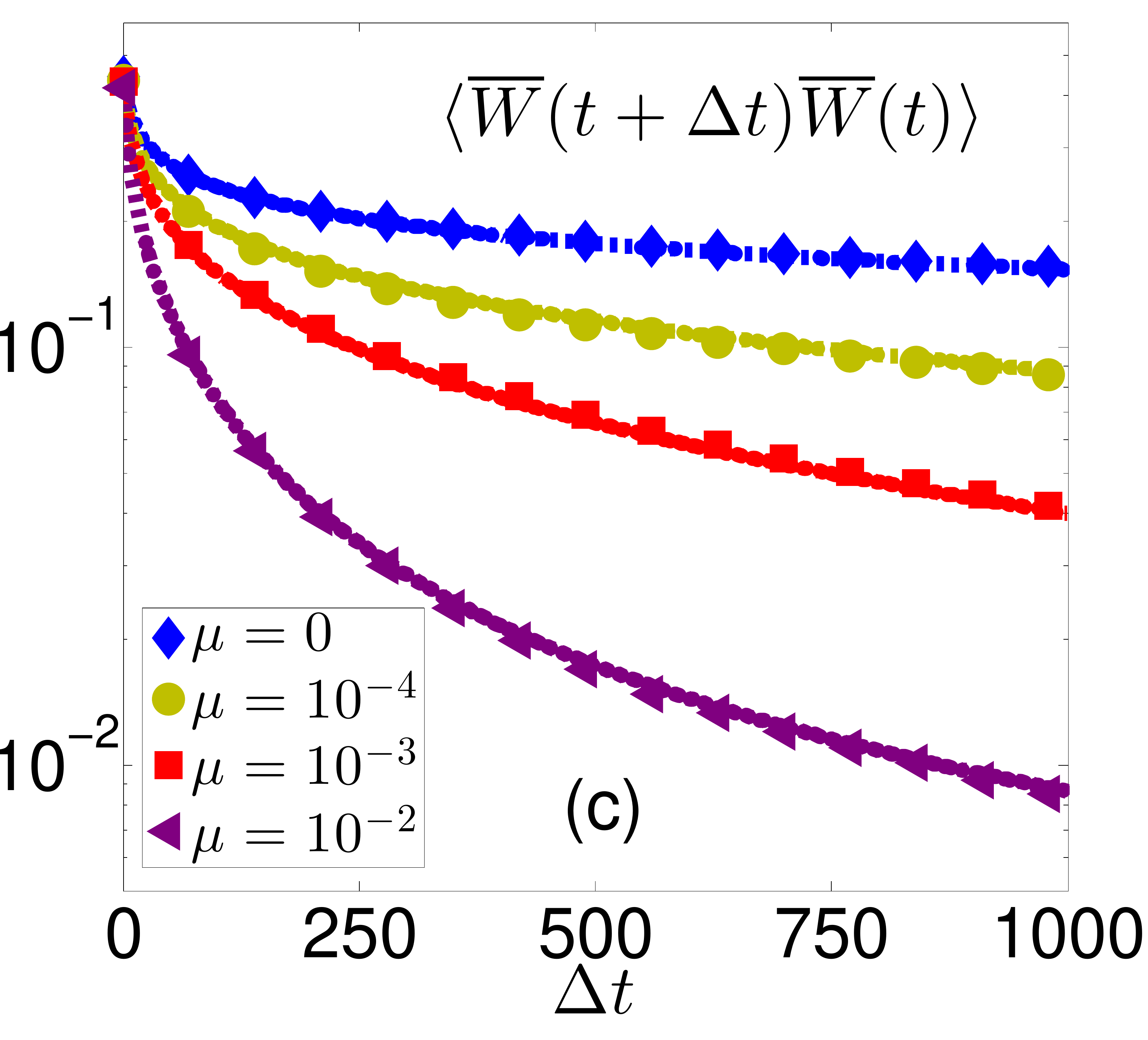}}
\subfloat[][]{\includegraphics[scale=0.135]{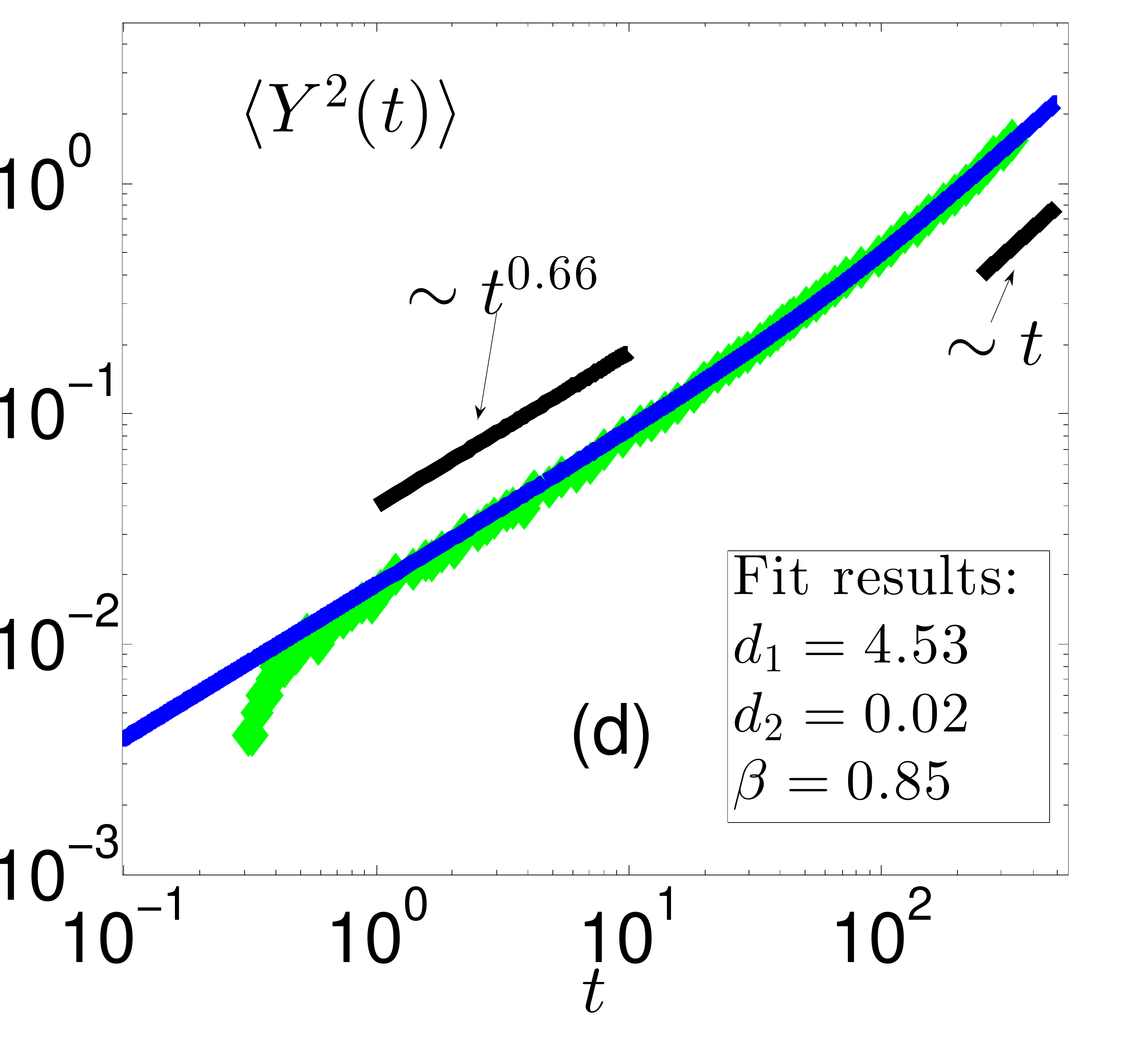}}\\[-4ex]
\vspace{-0.2cm}
\caption{(Colors online) (a--c) Here, $X(s)$ is an Ornstein-Uhlenbeck process ($\gamma=1$, $\sigma=1$), $\eta(s)$ is a tempered L{\'e}vy stable noise ($\alpha=0.25$) and $U(x)=x$. We show: (a) MSD of $\overline{W}(t)=W(t)/t$ (for initial position $x_0=0$); (b,c) two-point correlation functions of $Y$ and $\overline{W}$ (for $x_0=\sigma/\gamma$ and finite $t$). (d) Fit of the MSD data of mitochondria diffusing in \textit{S.~cerevisiae} cells depleted of actin microfilaments of Ref.~\cite{senning2010actin}. Here, $X(s)$ is pure diffusion ($F(x)=0, \, \sigma(x)=1$) and $\Phi$ is given by Eq.~(\ref{phi2}) with $\alpha_1=1$, $\alpha_2=0.66$.
}\label{fig:2} 
\vspace{-0.5cm}
\end{figure}

As specific example we consider $\eta(s)$ as a tempered L\'evy-stable noise with tempering index $\mu$ and stability index $0 < \alpha \leq 1$ interpolating between exponentially distributed ($\mu\to\infty$) and power-law distributed ($\mu=0$) waiting times \cite{stanislavsky2008Diffusion,baeumer2010tempered,gajda2010fractional,stanislavsky2014anomalous}. This implies that $\Phi(\lambda)=\left(\mu+\lambda\right)^{\alpha}-\mu^{\alpha}$ and thus $K(t)=e^{-\mu t} t^{\alpha-1} \MittagLeffler{\alpha}{\alpha}{(\mu t)^{\alpha}}$ \cite{janczura2011anomalous}.
We consider the case when $X(s)$ is given as an Ornstein-Uhlenbeck process $( F(x)=-\gamma\,x, \, \sigma(x)=\sqrt{2\,\sigma}$ in Eq.~\eqref{eq:1.1a}), such that $Y(t)$ intermediates between a CTRW and a normal diffusive oscillator. The MSD of the time-averaged $Y(t)$-process as a function of time exhibits an $\alpha$-dependent plateau for $t\to\infty$ in the CTRW limit ($\mu=0$) highlighting the ergodicity breaking of the process \cite{turgeman2009fractional}. For $\mu\neq 0$ we see that the MSD shows the CTRW scaling for short times, but converges to zero for $t\to\infty$ as in the Brownian limit confirming the ergodic nature of this anomalous process (Fig.~\ref{fig:2}a). This highlights that the MSD needs to be observed for a sufficiently long time to properly assess ergodicity breaking. We also obtain the associated two-point correlation functions.
The effect of $\mu\neq 0$ is clearly visible (Fig.~\ref{fig:2}b,c), which allows to distinguish between a CTRW and a process with waiting times distributed according to a tempered L\'evy-stable law. Remarkably, the functions $f_i(t)$ with $i=1,2$ are given in analytical form in this case:
\begin{subequations}
\begin{align}
f_1(t)&=\frac{\sigma}{\gamma(\gamma-\mu^{\alpha})}\left[-\mu^{\alpha}+ \gamma g(\alpha,\gamma,\mu;t)\right], \label{eq:3.8a} \\
f_2(t)&=\frac{\sigma}{\gamma}\left[\frac{1}{\gamma-\mu^{\alpha}}\left(\mu^{2 \alpha} - \gamma^2 g(\alpha,\gamma,\mu;t)\right)\right. \label{eq:3.8b} \\ 
& \quad \left. +\frac{1}{\Gamma(1-\alpha)}t^{-\alpha}e^{-\mu t}+\frac{\mu^{\alpha}}{\Gamma(1-\alpha)}\gamma(1-\alpha;\mu t)\right], \notag
\end{align}
\end{subequations}
where $\gamma(a;x)=\int_{0}^{x}t^{\alpha-1}e^{-t}\diff{t}$ is the incomplete gamma function and we define the new function $g(\alpha,\gamma,\mu;t)$ as an infinite series of confluent Kummer functions: 
\begin{equation}
g(\alpha,\gamma,\mu;t)=\sum_{n=0}^{\infty}\frac{(-1)^{n}((\gamma-\mu^{\alpha})t^{\alpha})^n}{\Gamma(1+\alpha n)}M(\alpha n, 1+\alpha n,-\mu t) \label{eq:3.9}
\end{equation}

We now apply our formalism to MSD data exhibiting crossover scaling between subdiffusive and normal diffusive regimes, as it is frequently observed in experiments \cite{bronstein2009transient,senning2010actin,jeon2012anomalous}. Fig.~\ref{fig:2}d shows the MSD of mitochondria diffusing in mating \textit{S. cerevisiae} cells, depleted of actin microfilaments, obtained with Fourier imaging correlation spectroscopy \cite{senning2010actin}. The crossover from a transient subdiffusive scaling with $\alpha=0.66$ to normal diffusion can not be captured quantitatively by the tempered L\'evy-stable $\Phi$, since the curvature at the crossover between the two pure power laws can not be modulated. We suggest instead a more flexible double power law form:
\begin{equation}
\Phi(\lambda)=d_1\left(\frac{\lambda}{d_2}\right)^{\alpha_1}\left(1+\left(\frac{\lambda}{d_2}\right)^{1/\beta}\right)^{(\alpha_2-\alpha_1)\beta},
\label{phi2}
\end{equation}
interpolating between power laws with exponents $\alpha_{1,2}$, with curvature tuned by the parameter $\beta$. Using a purely diffusive $X(s)$ process together with a least-squares method to determine the parameters of Eq.~(\ref{phi2}) yields an excellent model of the experimental data across the double power-law region. With the appropriate $\Phi$ our results immediately predict the quantitative form of the higher-order correlation functions of the diffusion process and its observables, which can be readily tested. Our framework also allows for a straightforward simulation of the underlying diffusion process by implementing the coupled Langevin Eqs.~(\ref{eq:1.1a},\ref{eq:1.1b}). In this way, one can predict many other quantities of interest, e.g., first passage time statistics, providing further testable predictions of the anomalous model with generalized waiting times.




\begin{acknowledgements}          

This research utilized Queen Mary's MidPlus computational facilities, supported by QMUL Research-IT and funded by EPSRC grant EP/K000128/1. We thank R. Klages for helpful discussions and A. H. Marcus for providing the data in Fig.~\ref{fig:2}d.

\end{acknowledgements}


%

\end{document}